\documentclass[twoside,slac_one]{revtex4}
\usepackage{graphicx}
\usepackage{fancyhdr}
\usepackage{amsmath} 
\usepackage{bm}
\usepackage{amsxtra}
\usepackage{amssymb}
\usepackage{amsthm}
\usepackage{latexsym}
\usepackage{lscape}

\newcommand{\NGEN}{N_{\mathrm{gen}}}
\newcommand{\NACC}{N_{\mathrm{acc}}}
\newcommand{\NEFF}{N_{\epsilon}}
\newcommand{\MLL}{M(\ell\ell)}

\newcommand{\rhoeff}{\rho_{\mathrm{eff}}}
\newcommand{\effdata}{\varepsilon_{\mathrm{data}}}
\newcommand{\effmc}{\varepsilon_{\mathrm{sim}}}

\newcommand{\effreco}{\varepsilon_{\mathrm{reco}}}
\newcommand{\effid}{\varepsilon_{\mathrm{id}}}
\newcommand{\efftrig}{\varepsilon_{\mathrm{trig}}}

\newcommand{\ANORM}{A_{\mathrm{norm}}}
\newcommand{\LUM}{{\cal{L}}}
\newcommand{\NUNORM}{N_{u,{\mathrm{norm}}}}
\newcommand{\effNORM}{\varepsilon_{\mathrm{norm}}}
\newcommand{\rhoNORM}{\rho_{\mathrm{norm}}}

\newcommand{\RPOSTFSR}{R_{\text{post-FSR}}}
\newcommand{\RPOSTFSRDET}{R_{\text{det,post-FSR}}}
\newcommand{\RDET}{R_{\mathrm{det}}}

\def\pt{\ensuremath{\mathrm{p_T}}}
\def\PT{\ensuremath{\mathrm{p_T}}}
\def\ET{\ensuremath{\mathrm{E_T}}}
\def\GeV{\ensuremath{\mathrm{GeV}}}
\def\TeV{\ensuremath{\mathrm{TeV}}}
\def\ee{\ensuremath{\mathrm{e^+e^-}}}
\def\pbinv{\ensuremath{\mathrm{pb^{-1}}}}

\begin{document}

\title{Measurement of the Drell--Yan
differential cross section at 7 TeV }

%

\author{S. Stoynev {\it{on behalf of the CMS Collaboration}}}
\affiliation{Department of Physics and Astronomy, Northwestern University, Evanston, IL, USA}

\begin{abstract}
  The Drell--Yan differential cross section is measured in
  $\ensuremath{\mathrm{pp}}$  collisions at   $\sqrt{s} = 7$~\TeV,
  from a data sample
  collected with the CMS detector at the LHC, corresponding to an integrated luminosity of $36~{\mathrm{pb}}^{-1}$.
  The cross section measurement, normalized to the measured cross section in the Z region, is reported for both the dimuon and dielectron channels in the
  dilepton invariant mass range 15--600 GeV.
  The normalized cross section values are quoted both
  in the full phase space and within the detector acceptance. The effect of final state
  radiation is also identified. The results are found to agree with theoretical predictions.

\end{abstract}

\maketitle

\thispagestyle{fancy}

\section{Introduction}

The production of lepton pairs in hadron-hadron collisions via the Drell--Yan (DY) process
is described in the standard model (SM) by the $s$-channel exchange of $\gamma^*/\text{Z}$.  
Theoretical calculations of the differential cross section $d\sigma/d\MLL$, where $\MLL$ is the 
dilepton invariant mass, are well established up to 
next-to-next-to-leading order (NNLO)~\cite{DY-theory, DYNNLO, DYNNLO1}.
Therefore, comparisons between calculations and precise experimental measurements
provide stringent tests of perturbative quantum chromodynamics (QCD) and significant 
constraints on the evaluation of the parton distribution functions (PDFs).
Furthermore, the production of DY lepton pairs constitutes a major source of background
for top quark pair
and diboson measurements, as well as for searches for new physics, such as 
production of high mass dilepton resonances.

Here we present a measurement of the differential DY cross section
in proton-proton collisions at $\sqrt{s} = 7$~\TeV, based on dimuon and dielectron 
data samples collected in 2010 by the Compact Muon Solenoid (CMS) experiment at the Large
Hadron Collider (LHC), corresponding to an integrated luminosity of $35.9\pm 1.4$~\pbinv.
The results are given for the dilepton invariant mass range $15 < \MLL < 600$~\GeV.
In the analysis presented, the cross sections are calculated as 

\begin{equation}\label{eqn:fullCrossSection_intro}
\sigma = \frac{N_{\text{u}}}{A \, \epsilon \, \rho \, \LUM}\, , 
\end{equation}

where $N_{\text {u}}$ is the unfolded background-subtracted yield, corrected for detector 
resolution.  The values of the acceptance $A$ and the efficiency $\epsilon$ are 
estimated from simulation, while $\rho$ is a factor that accounts for differences 
in the detection efficiency between data and simulation.  
Knowledge of the integrated luminosity $\LUM$ is
not required for the measurements described in this paper, since the cross
sections are normalized to the Z region ($60 < \MLL < 120$~\GeV). 
The measurements are described in detail in Ref.~\cite{DY_CMS}.


\section{The CMS Detector}
A detailed description of the CMS detector and its performance can be found
in Ref.~\cite{ref:CMS}. The central feature of the CMS apparatus is a
superconducting solenoid 13~m in length and 6~m in diameter, which
provides an axial magnetic field of 3.8~T. Within the field volume are
the silicon pixel and strip tracker, the crystal electromagnetic
calorimeter (ECAL), and the brass/scintillator hadron calorimeter (HCAL). 
Charged particle trajectories are measured by the
tracker, covering 
 the full azimuthal angle
and pseudorapidity interval $|\eta| < 2.5$, 
where the pseudorapidity is defined as 
$\eta = -\ln \tan (\theta/2)$,
with $\theta$ being the polar angle of the trajectory of the particle
with respect to the counterclockwise beam direction.  
Muons are measured in the pseudorapidity range $|\eta|< 2.4$. 
\section{Event Selection}

The analysis presented in this paper is based on dilepton data samples
selected by inclusive single-lepton triggers with \pt~thresholds~
ranging between 9 and 15~\GeV~ for muons and between 15 and 17~\GeV~ 
for electrons, depending on the beam conditions. 

Muons are required to pass the standard CMS muon
identification and quality criteria, based on the number of hits found
in the tracker, the response of the muon chambers, and a set of
matching criteria between the muon track parameters as determined by
the inner tracker section of the detector and as measured in the muon 
chambers~\cite{tag-and-probe, ref:CRAFT}. 
Cosmic-ray muons are further suppressed by requirements on the
proximity to the interaction region and the opening angle between the two 
reconstructed muons. The two muons are fitted to a common vertex 
reducing the contamination from QCD related events.
Finally, an isolation requirement is imposed on each muon, based on the
relative transverse energy deposited in the HCAL and transverse 
momenta of tracks around each of the muons.  

Electron reconstruction starts from clusters of energy deposited in
the ECAL, and associates with them hits
in the CMS 
tracker~\cite{EGMPAS}. 
Energy-scale corrections are
applied to individual electrons as described in Ref.~\cite{WAsymmetry}. 
The electron candidate is required to be consistent with 
a particle originating from the primary vertex in the event. Electron 
identification criteria based on shower shape and track-cluster 
matching are applied to the reconstructed candidates.
Electrons originating from photon conversions are rejected by 
eliminating those electrons for which a partner track consistent with
a conversion hypothesis is found, and requiring no missing hits
in the pixel detector,
as discussed in Ref.~\cite{ZCrossSection}.
Isolation requirements are imposed on each electron, 
based on the relative transverse energy deposited in the ECAL and the HCAL and 
transverse momenta of tracks around each of the muons.

Muons must be reconstructed with $|\eta|<2.4$ and $\PT > 7$~\GeV~ but at
 least one of them is required to be with $|\eta|<2.1$ and $\PT > 16$~\GeV~
in order to ensure effective triggering of the event.
Electrons must be reconstructed in the ECAL barrel
with $|\eta| < 1.44$ or in the ECAL endcaps with
$1.57 < |\eta| < 2.5$.
The leading electron is required to have $\ET > 20$~\GeV~
providing it triggered the event, 
while the second electron must have $\ET > 10$~\GeV.

Event samples for simulation studies of electroweak
processes involving W and Z production are produced with the NLO 
MC
generator
{\sc POWHEG}~\cite{Alioli:2008gx, Nason:2004rx, Frixione:2007vw} interfaced
with the {\sc PYTHIA} (v.~6.422)~\cite{Sjostrand:2006za} parton-shower event generator, using
the CT10~\cite{CT10} parametrization of the PDFs. {\sc PYTHIA} is also used for the FSR simulation. 
The QCD multijet background is
generated with {\sc PYTHIA}, and  
background from top quark pairs is simulated using {\sc MadGraph} (v.~4.4.12)~\cite{MadGraph} and {\sc PYTHIA}
at leading order using the CTEQ\,6L PDF set~\cite{CTEQ} for both samples. 
Generated events are processed through the full {\sc GEANT4}~\cite{GEANT4}
detector simulation, trigger emulation, and event reconstruction chain.

\begin{figure}[t!]
  \begin{center}
    \includegraphics[width=0.40\textwidth, angle=90]{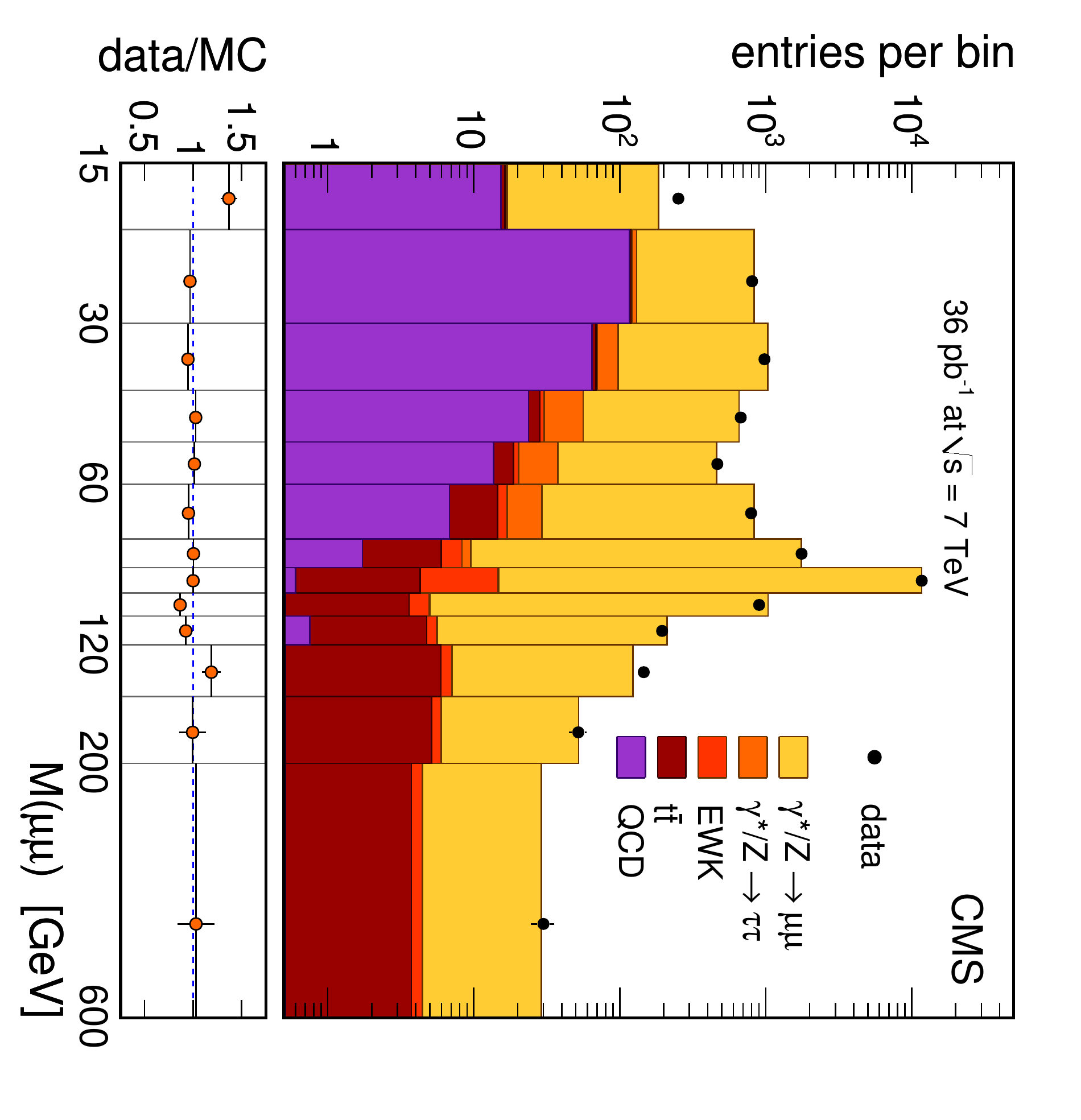}
    \includegraphics[width=0.40\textwidth, angle=90]{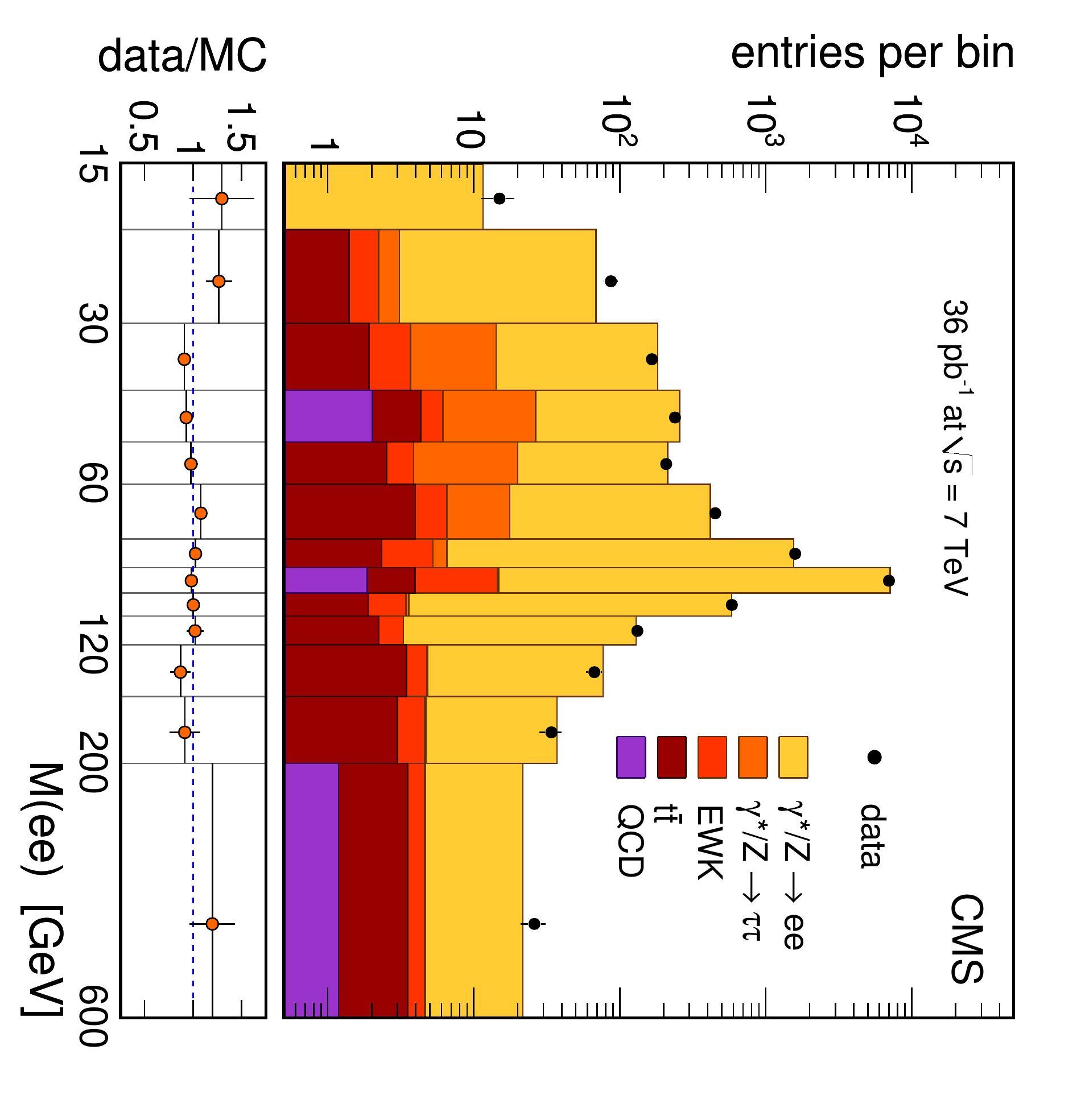}
    \caption{The observed dimuon (left) and dielectron (right) invariant mass spectra. No corrections are applied to the distributions. The points with error bars represent the data, while the 
various contributions from simulated events are shown as
stacked histograms. By ``EWK'' we denote $\text{W} \rightarrow \ell\nu$ and diboson production. The ``QCD'' 
contribution results from processes associated with QCD and could be genuine or misidentified leptons.  
The lower panels show the ratios between the measured and the 
simulated distributions including the statistical uncertainties from both.
       \label{fig:mass-observed}
    }
  \end{center}
\end{figure}

The observed invariant mass distributions, in the dimuon and dielectron channels, are shown in 
Fig.
~\ref{fig:mass-observed}. 
Thirteen mass bins with unequal widths are used to cover the observable dilepton mass
spectrum.             

\section{Backgrounds}
The main backgrounds at 
high dilepton invariant masses are caused by top quark pairs
and diboson production, while at  
invariant masses below the $Z$ peak, DY production of $\tau^+\tau^-$ pairs 
becomes the dominant background. At low dimuon invariant masses, most background events 
are QCD multijet events that 
 originate from both heavy flavor and light quark production.
The expected shapes and 
relative yields of these several dilepton sources can be seen in 
Fig.~\ref{fig:mass-observed}. They are in agreement with the estimates extracted from data
and the significant background sources are estimated from data\cite{DY_CMS}. 

\section{Detector Resolution Effects and Unfolding}

The effects of the detector
resolution on the observed dilepton spectra are corrected through an
unfolding procedure. The original invariant mass spectrum is
related to the observed one (in the limit of no background) by
\begin{equation}
   N_{{\mathrm{obs},i}} = \sum_k \, T_{ik} \, N_{\mathrm{true},k} ,
\end{equation}
where $N_i$ is the event count in a given invariant mass bin~$i$. 
The element $T_{ik}$ of the ``response matrix'' $T$ is the probability 
that an event with an original invariant mass in the bin $k$ is reconstructed 
with an invariant mass in the bin $i$. 
The original invariant mass spectrum is obtained by inverting the response 
matrix and calculating~\cite{Cowan-unfolding,Bohm-unfolding} 
\begin{equation}
     N_{\text{u},k} \equiv N_{\mathrm{true},k} = \sum_i \, (T^{-1})_{ki} \, N_{{\mathrm{obs}},i} .
\label{eq:invResponse}
\end{equation}
This procedure is sufficient in the analysis reported in
this paper 
because the response matrix is nonsingular and nearly diagonal. 
Two extra dilepton invariant mass bins are included in the unfolding procedure, 
to account for events observed with $\MLL < 15$~\GeV~ or 
$\MLL > 600$~\GeV.

\begin{figure}[hbtp]
  \begin{center}
    \includegraphics[width=0.40\textwidth]{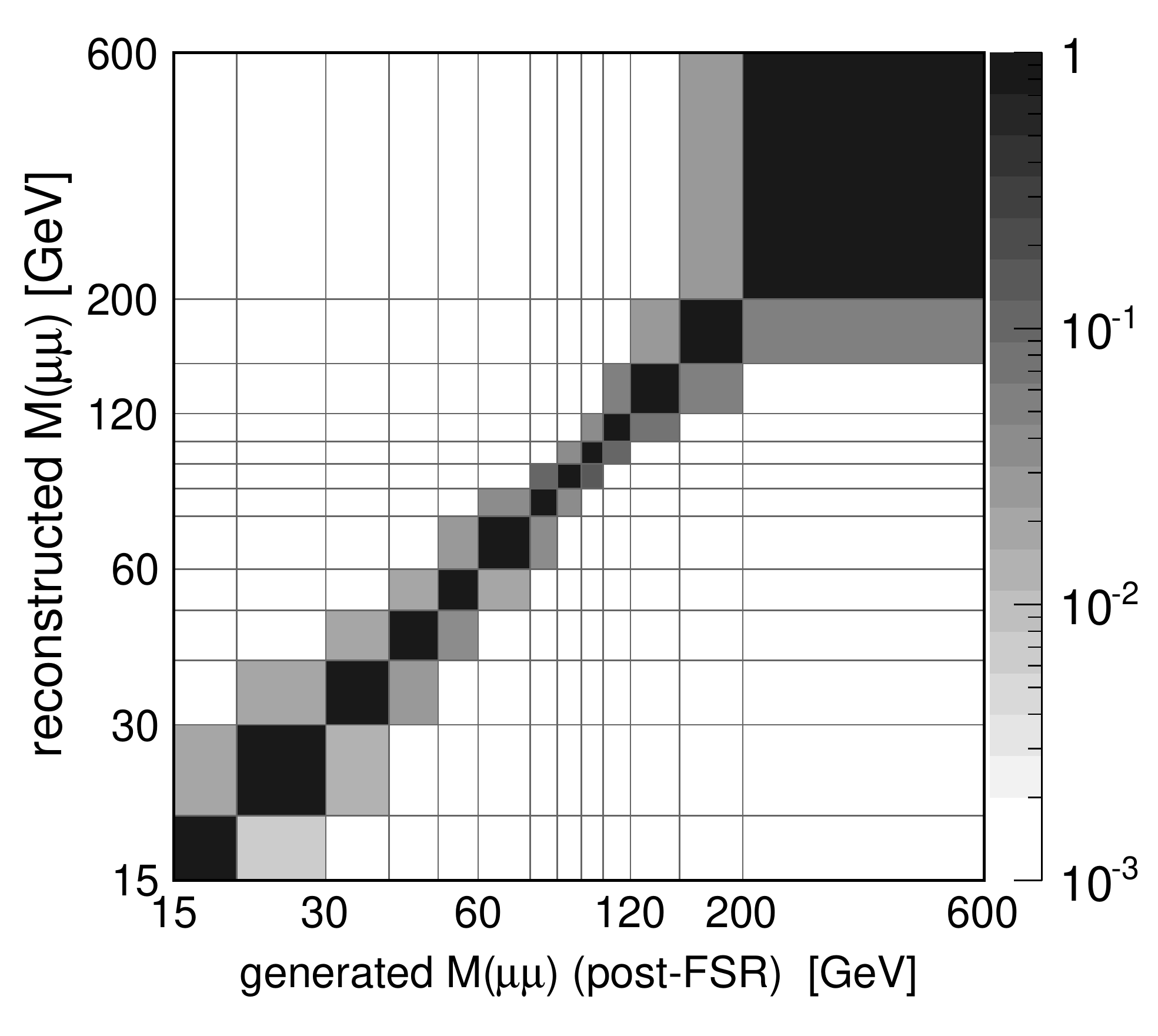}
    \includegraphics[width=0.40\textwidth]{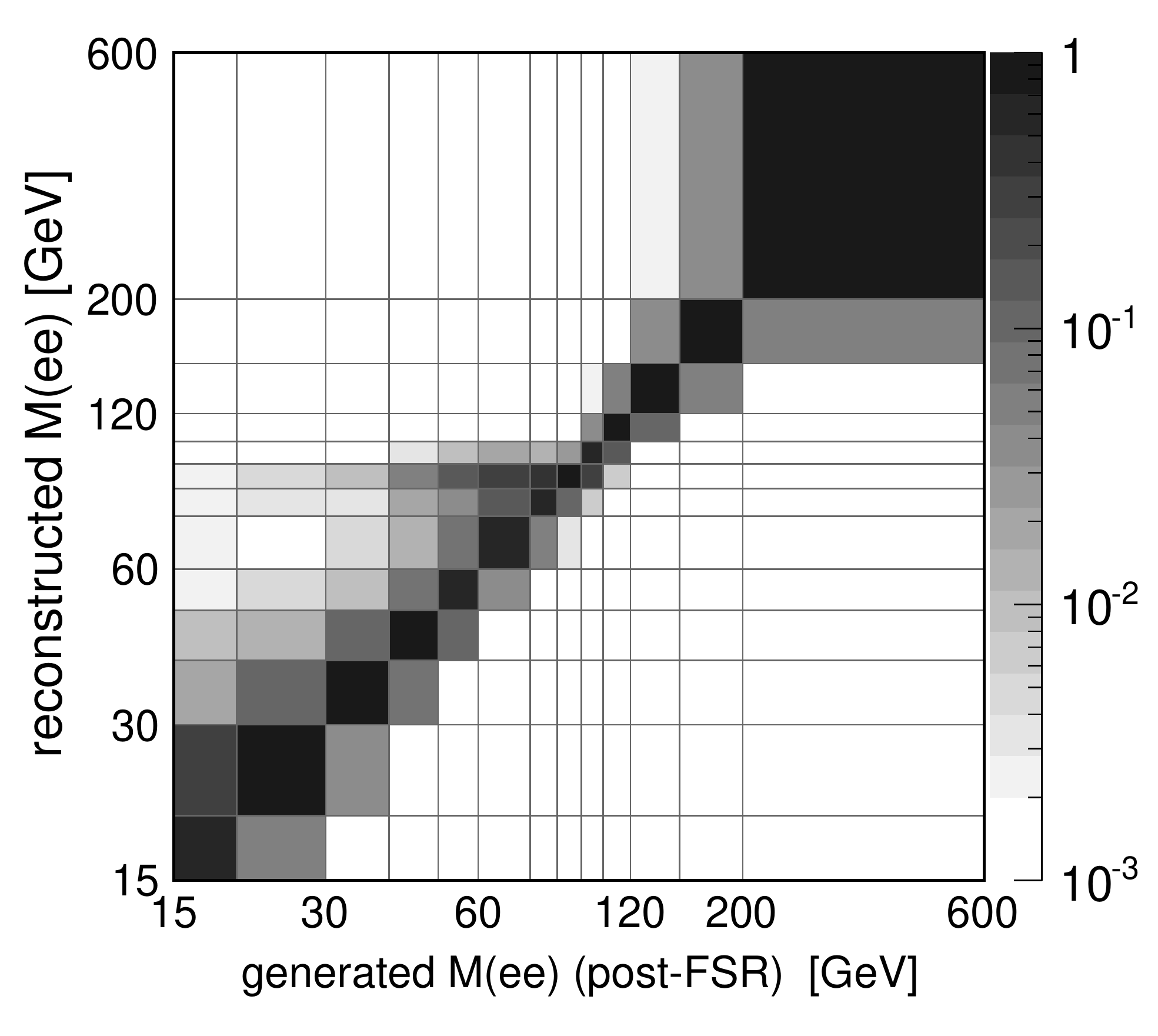}
    \caption{The response matrices for the muon (left) and electron (right) 
      channels from simulation.
       \label{fig:response-matrix}
    }
  \end{center}
\end{figure}

The response matrix is calculated using the simulated sample of DY 
events, defining the ``true mass'' as the ``generator level'' dilepton invariant mass, 
after 
FSR. Only the 
selected
events in the sample are used to calculate the response matrix. The loss of events
caused by reconstruction inefficiencies or limited acceptance is
factored out from the unfolding procedure and taken into account
by means of efficiency and acceptance factors in a 
subsequent step. 
The response matrices in
both lepton channels are invertible and are represented on Fig.~\ref{fig:response-matrix}.

\section{Acceptance and Efficiency}

The geometrical and kinematic acceptance $A$ is defined, using the simulated leptons after the
FSR simulation, as $A \equiv \NACC/\NGEN$, where $\NGEN$ is the number of generated events and
$\NACC$ is the corresponding number of events passing the standard $\PT$ and $\eta$ 
lepton requirements, in each dilepton invariant mass bin.

The efficiency $\epsilon$ is the fraction of events within the acceptance that pass the full 
selection, so that
\begin{equation}\label{eqn:AccEff}
    A \cdot \epsilon \equiv \frac{\NACC}{\NGEN} \cdot
    \frac{\NEFF}{\NACC} = \frac{\NEFF}{\NGEN} ,
\end{equation}
where $\NEFF$ is the number of events surviving the reconstruction, selection, and identification requirements.
The values of the product of acceptance and efficiency are obtained from simulation. 
A separate 
correction factor is determined from data and applied to the product, following the procedure used in the inclusive W 
and Z cross section measurements in CMS~\cite{ZCrossSection}. This factor, the efficiency correction,
describes the difference 
between data and simulation
in the efficiency to observe single leptons or dileptons.

The {\sc POWHEG} simulation combines the next-to-leading-order (NLO) calculations with a parton showering which 
is insufficient to model fully the low invariant mass region of the dilepton spectra.
The two high-\pt~ leptons required in the analysis must form a small angle at low mass and therefore the dilepton 
system gets significantly boosted, something to be compensated by hard gluon radiation in the transverse plane. This means 
that these low-mass events are of the type ``$\gamma^*$ + hard jet'' at first order, and therefore the next order of 
correction (NNLO) becomes essential for a reliable estimate of acceptance corrections.
To account for this, a
correction is applied, determined from the ratio between the differential cross sections calculated
at NNLO with {\sc FEWZ}~\cite{FEWZ} and at NLO with {\sc POWHEG}, both before FSR.  These correction weights, obtained
in bins of dilepton rapidity, \pt, and invariant mass, are applied on an event-by-event basis.
This procedure changes the acceptance in the lowest invariant mass bin significantly (by about 50\%), 
but has a small effect, not exceeding 3\%, on the rest of the bins.

\begin{figure}[h]
{\centering
\includegraphics[width=0.30\textwidth, angle = 90]{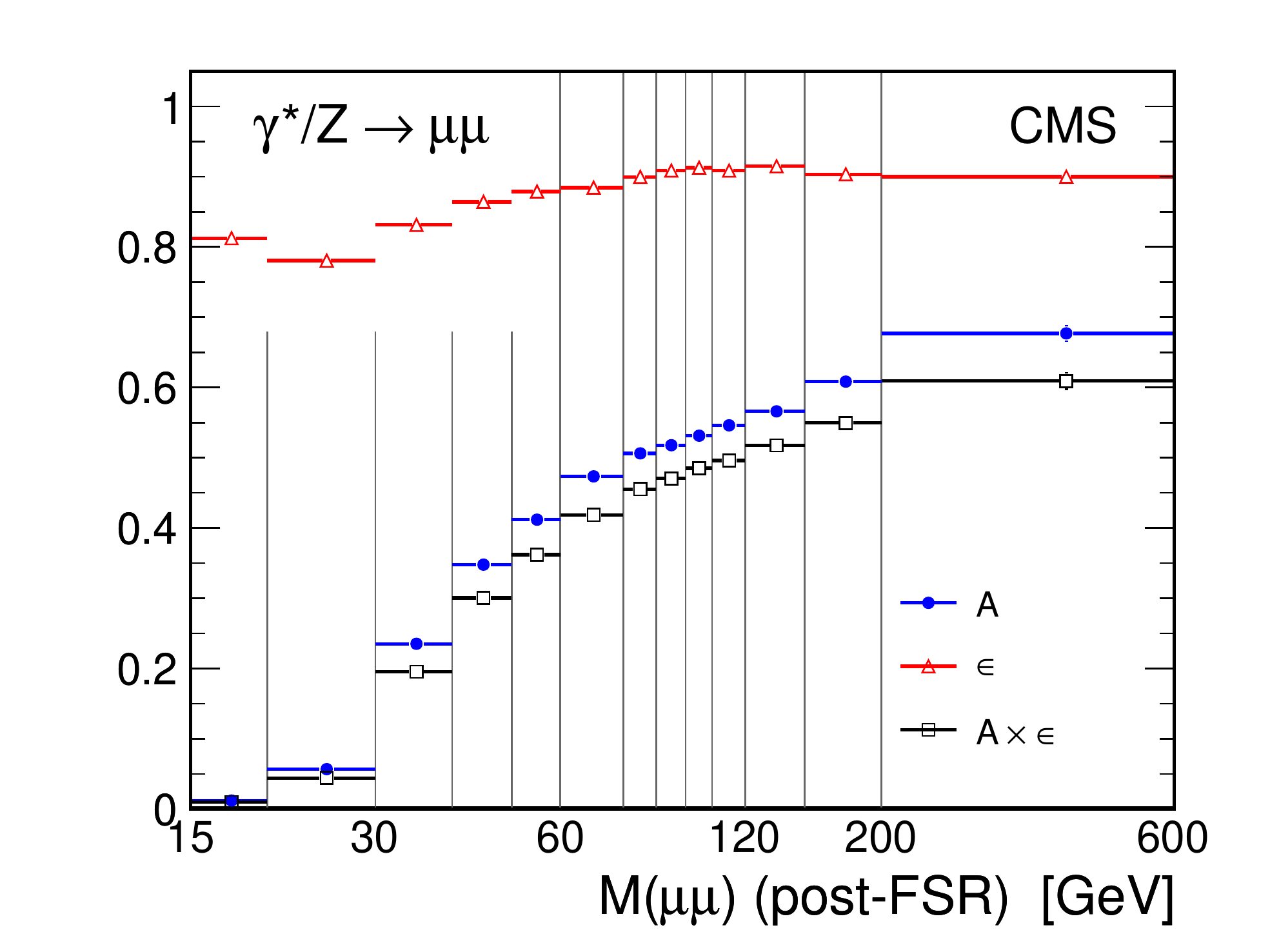}
\includegraphics[width=0.30\textwidth, angle = 90]{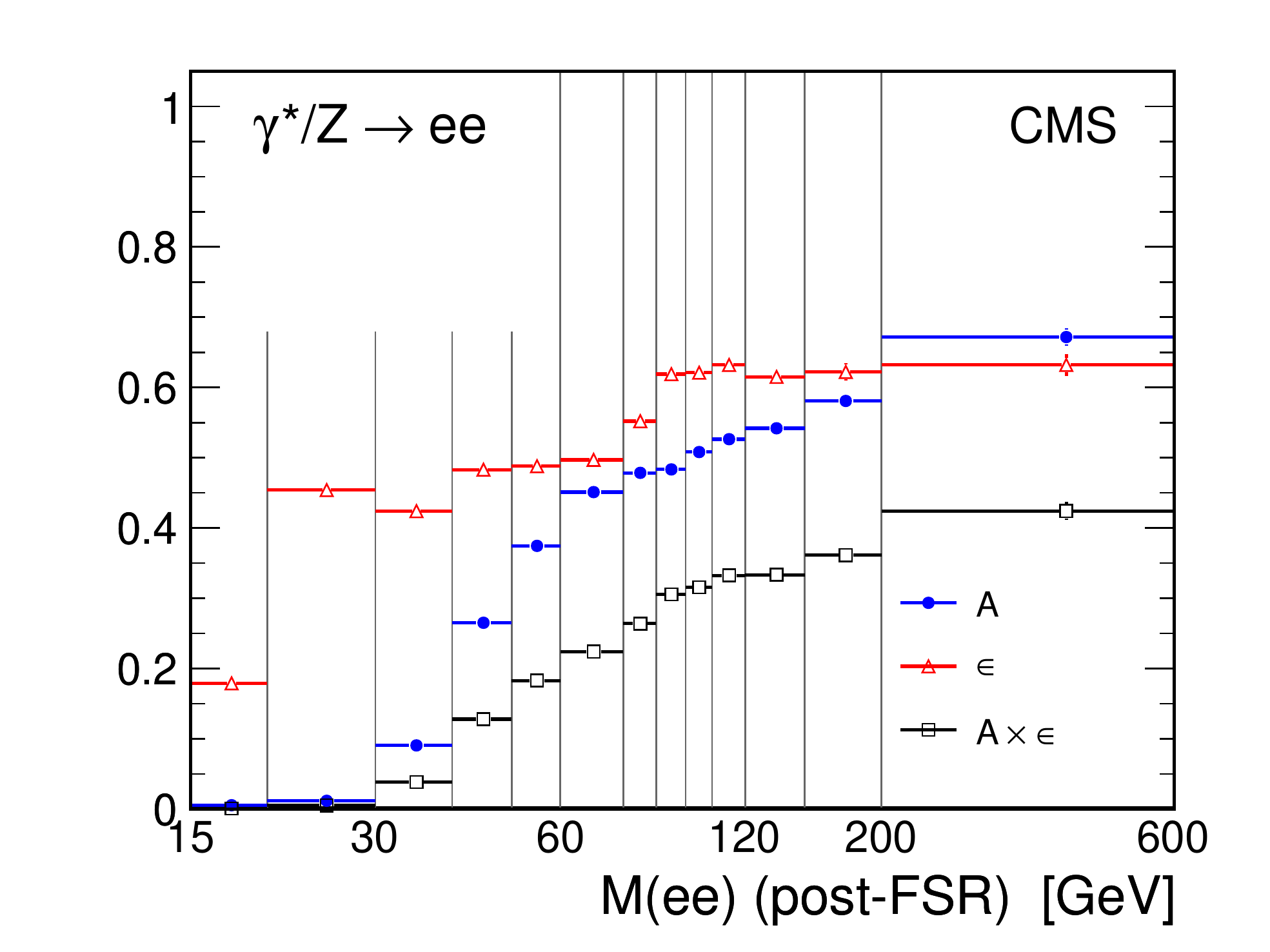}
\caption{\label{Acc}
DY acceptance (blue, filled circles), efficiency (red, open triangles), and their product (black, open squares) per invariant mass bin, 
for the $\mu^+\mu^-$ (left) and $\ee$ (right) channels.}
}
\end{figure}

Figure~\ref{Acc} shows the variables $A$, $\epsilon$, and
$A\cdot\epsilon$ as functions of $\MLL$ for dimuons (left) and
dielectrons (right).

The FSR correction factors 
for
a given invariant mass range are obtained 
from simulation by dividing the cross sections after FSR by the corresponding 
quantities before FSR. The factors are channel and invariant mass dependent and are up to
few hundred of percent just below the Z peak but  ~10-20 \% or less elsewhere. They are applied on 
(corrected) data as an additional step.
The factors obtained within the detector acceptance and in the full phase space 
are applied to the corresponding measurements.

The total dimuon event selection efficiency is factorized as

\begin{equation}
\varepsilon(\text{event}) = 
       \varepsilon(\mu_1)
       \cdot \varepsilon(\mu_2)
       \cdot \varepsilon[\mu\mu|(\mu_1) \& (\mu_2)]
       \cdot \varepsilon(\text{event},\text{trig}|\mu\mu) ,
\label{eq:muon-event-eff}
\end{equation}

where $\varepsilon(\mu)$ is the single muon selection efficiency;
$\varepsilon[\mu\mu|(\mu_1) \& (\mu_2)]$ is the dimuon 
selection efficiency, which includes 
the requirement that 
the two muon tracks be consistent with originating from a common vertex and that they satisfy the angular criteria;
and
$\varepsilon(\text{event},\text{trig}|\mu\mu)$ is the efficiency of triggering an event.
The single muon efficiency is factorized as

\begin{equation}
\varepsilon(\mu)= \varepsilon(\text{track}|\text{accepted})
                             \cdot\varepsilon(\text{reco}+\text{id}|\text{track})
                             \cdot\varepsilon(\text{iso}|\text{reco}+\text{id}) ,
\label{eq:single-muon-eff}
\end{equation}

where 
$\varepsilon(\text{track}|\text{accepted})$ is the offline track reconstruction efficiency in the tracker detector;
$\varepsilon(\text{reco}+\text{id}|\text{track})$ is the muon reconstruction and identification efficiency; and
$\varepsilon(\text{iso}|\text{reco}+\text{id})$ is the muon isolation efficiency.
The trigger efficiency $\varepsilon(\text{event},\text{trig}|\mu\mu)$ is given by

\begin{equation}
\varepsilon(\text{event},\text{trig}|\mu\mu) 
     = \varepsilon(\mu_1,\text{trig}|\mu_1) 
     + \varepsilon(\mu_2,\text{trig}|\mu_2)
     - \varepsilon(\mu_1,\text{trig}|\mu_1)
         \cdot\varepsilon(\mu_2,\text{trig}|\mu_2) ,
\end{equation}

where $\varepsilon(\mu,\text{trig}|\mu)$ 
is the efficiency of an offline selected muon to fire the trigger.

The track reconstruction efficiency is very high ($99.5\%$). The angular 
criterion is nearly $100\%$ efficient for signal DY events, and the vertex 
probability requirement is more than $98\%$ efficient and has a negligible 
($< 0.3\%$) dependence on~$\MLL$.

The muon reconstruction and identification efficiency is estimated using 
clean samples of muon pairs in the Z peak (tag and probe, T\&P, method~\cite{ZCrossSection}). 
To determine the isolation efficiency, the Lepton Kinematic Template Cones (LKTC) method~\cite{tag-and-probe}  is applied.

To describe the observed efficiency variations between data and simulation, efficiency correction factors are obtained in bins of $\PT$ and $\eta$
as the ratio of the efficiencies measured with
data and with the simulated events:
\begin{equation}
\rhoeff(\PT,\eta) = \frac{\effdata(\PT,\eta)}{\effmc(\PT,\eta)} .
\label{eqn:rho-definition}
\end{equation}

The corrections to the efficiencies in simulation are implemented by reweighting simulated events, with 
weights computed as 
$W =\rho_1^{\text{reco+id}}  \rho_2^{\text{reco+id}}    \rho_1^{\text{iso}}   \rho_2^{\text{iso}} \rho^{\text{trig}}$ where 
$\rho^X_i$ is estimated per muon~$i=1, 2$  from Eq.~(\ref{eqn:rho-definition}) 
using $X="reco+id", "iso"$ efficiencies defined in Eq.~(\ref{eq:single-muon-eff})  and
$\rho^{\text{trig}} = (\varepsilon_{\text{data},1}^{\text{trig}} +  \varepsilon_{\text{data},2}^{\text{trig}} - \varepsilon_{\text{data},1}^{\text{trig}}  \varepsilon_{\text{data},2}^{\text{trig}})/(\varepsilon_{\text{sim},1}^{\text{trig}} +  \varepsilon_{\text{sim},2}^{\text{trig}} - \varepsilon_{\text{sim},1}^{\text{trig}}  \varepsilon_{\text{sim},2}^{\text{trig}})$.
If $\PT < 16$~\GeV~ or $|\eta| > 2.1$ for a given muon, its 
trigger efficiency
is set to zero.

The total event efficiency in the dielectron channel analysis is defined
as the product of the two single electron efficiencies, which incorporate three
factors: 1)~the efficiency $\effreco$ to reconstruct an electron candidate from an energy
deposit in the ECAL; 2)~the efficiency $\effid$ for that candidate to pass the selection
criteria, including identification, isolation, and conversion rejection; 3)~the efficiency
$\efftrig$ for the leading electron to pass the trigger
requirements. Each of these efficiencies is obtained from simulation and corrected by
$\rhoeff(\PT,\eta)$, as for the muon channel (Eq.~(\ref{eqn:rho-definition})).  The 
T\&P method is used for all efficiency components.  The event efficiency correction
and its uncertainty are derived as for the muon channel by reweighting simulated events.

The overall efficiency correction is up to 10 percent with largest values at lowest invariant masses.


\section{Systematic Uncertainties}

The acceptance-related uncertainty resulting from the knowledge of the PDFs is estimated using {\sc PYTHIA} 
with the CTEQ6.1 PDF set by a reweighting technique~\cite{Bourilkov:2006cj}. 
It is approximately the same for the dimuon and dielectron channels and is between one and three percent
in the whole invariant mass range under study.  

The uncertainty of the acceptance is estimated, for each dilepton invariant mass bin,
using {\sc FEWZ}, at NLO and NNLO accuracy
in perturbative QCD.   Variations of the factorization and renormalization
scales lead to a systematic uncertainty smaller than 1\% (at NNLO) 
for most of the invariant mass range used in the analysis presented here.

Since the {\sc POWHEG} MC (NLO) simulation, modified to match the {\sc FEWZ} (NNLO) calculations, is used 
to calculate the acceptance
corrections used in the analysis, 
an additional (model-dependent) systematic uncertainty on the acceptance
calculation is determined from the 
observed differences in acceptances based on
{\sc FEWZ} spectra and {\sc POWHEG} distributions matched to {\sc FEWZ}.
This systematic uncertainty reaches up to 10\% (at lowest invariant masses) in the dilepton
invariant mass range considered in the analysis and is included in the comparison 
between the measurements and the theoretical expectations.

The dominant systematic uncertainty on the cross section measurement in the 
dimuon channel is the uncertainty on the background estimation. It reaches four percent 
at very low invariant masses and is also substantial at high masses where the measurement 
is statistically limited. 
The next most important uncertainties are related to 
the muon efficiency and to the muon momentum scale and resolution, directly related to
the unfolding procedure.  Dimuons from the Z region 
and the invariant mass shape distribution are used to assess these uncertainties. They are of
the order of one percent except in the proximity to the Z peak (2-4 \%) and also at very high 
masses where statistics is poor (~2\%).   
Comparisons of the FSR spectrum between data and simulation show that potential differences 
could not affect the cross section measurements by more that 2\% and are typically confined 
within 0.5 \%.
Remaining systematic effects are within 1\% and/or not exeeding the statistical precision of the 
measurement.

In the electron channel, the leading systematic uncertainty is
associated with the couple of percent uncertainty on the energy scale correction of 
individual electrons. Its effect on the cross section measurement is very significant and reaches 
tens of percent at very low invariant masses and around the Z peak, it is several percent elsewhere. 
The second leading uncertainty for electrons is caused by the uncertainty on the
efficiency scale factors which is statistically limited. It is close to 10\%
in the low invariant mass region and few percent in the rest of the spectrum.
The dielectron background uncertainties follow the pattern described for muons 
but are slightly higher at low invariant masses reaching 4--6 \%. 
The resolution effects through the unfolding procedure are smaller than the leading 
source for each of the invariant mass bins and are typically few percent. 
Because of significantly higher
systematic uncertainty for all mass bins for the electron channel than
for the muon channel, the FSR related contribution to the electron channel
systematic uncertainty is neglected.


\section{Results}

The DY cross section per invariant mass bin $i$, $\sigma_i$, is calculated according to Eq. (\ref{eqn:fullCrossSection_intro}).

In order to provide a measurement independent of the luminosity uncertainty and to reduce
many systematic uncertainties, the $\sigma_i$ is normalized to
the cross section in the Z region, $\sigma_{\mathrm{\ell\ell}}$, defined as the DY 
cross section in the invariant mass region $60 < \MLL < 120~\GeV$.  
The result of the analysis is presented as the ratio
\begin{equation}
\label{eqn:fullCrossSectionRatio}
  R^i_{\text{post-FSR}} = \frac{N_{\text{u},i}}{A_i\,\varepsilon_i\,\rho_i} \big/
   \frac{\NUNORM}{\ANORM\,\effNORM\,\rhoNORM},
\end{equation}
where $N_{\text{u},i}$ is the number of events after the unfolding procedure, and the
acceptances $A_i$, the efficiencies $\epsilon_i$, and the corrections estimated from data,
$\rho_i$, were defined earlier; $\NUNORM$, $\ANORM$, $\effNORM$, and $\rhoNORM$ 
refer to the Z region.
For both lepton channels, the cross sections 
in the Z region measured in this analysis are in excellent agreement with the 
previous CMS measurement~\cite{ZCrossSection}.

In order to allow a more direct and precise comparison with theory predictions, the 
shape measured before the acceptance correction is also reported, thus eliminating PDF and theory 
uncertainties from the experimental results:
\begin{equation}
\label{eqn:fullCrossSectionRatio_DET}
  R_{\text{det, post-FSR}}^i = \frac{N_{\text{u},i}}{\varepsilon_i\,\rho_i} \big/
   \frac{\NUNORM}{\effNORM\,\rhoNORM} .
\end{equation}

The shapes corresponding to the DY process after FSR, $\RPOSTFSR$ and $\RPOSTFSRDET$, 
are modified by the FSR correction factors.
to obtain the shapes before FSR, $R$ and $\RDET$, respectively.
The shapes integrated in the normalization region are equal to one by construction. 

The results are presented in Tables~\ref{tab_result_single}
and~\ref{tab_result_electrons}, respectively, for the dimuon and dielectron
channels.
The two shape measurements, 
shown in the last column of the tables, 
are in good agreement for 11 out of 13 invariant
mass bins and remain statistically consistent (although marginally)
for the remaining two bins, 40--50~\GeV~ and 120--150~\GeV.

\begin{table}[h!]
\begin{center}
\caption{Results for the DY spectrum normalized to the Z region in the
dimuon channel. The statistical and systematic uncertainties
are summed in quadrature.   $\RPOSTFSR$ and $\RPOSTFSRDET$ are
calculated using Eqs.~(\ref{eqn:fullCrossSectionRatio}) 
and~(\ref{eqn:fullCrossSectionRatio_DET}), respectively. The $\RDET$ and $R$ are
calculated using the FSR corrections.
\label{tab_result_single} }
\begin{tabular}{|l|r@{$~\pm~$}l|r@{$~\pm~$}l|r@{$~\pm~$}l|r@{$~\pm~$}l|}
\hline
Invariant mass bin (\!\GeV) 
& \multicolumn{2}{c|}{$\RPOSTFSRDET~(10^{-3})$} 
& \multicolumn{2}{c|}{$\RDET~(10^{-3}) $}
& \multicolumn{2}{c|}{$\RPOSTFSR~(10^{-3}) $}
& \multicolumn{2}{c|}{$R~(10^{-3}) $}\\
\hline
15--20   & 18 & 2 & 19 & 2 & 772 & 67 & 780 & 69 \\
20--30   & 58 & 3 & 58 & 3 & 528 & 33 & 533 & 34 \\
30--40   & 67 & 3 & 67 & 3 & 147 & 8  & 147 & 8 \\
40--50   & 44 & 2 & 41 & 2 & 66  & 4  & 62  & 4 \\
50--60   & 30 & 2 & 23 & 2 & 37  & 3  & 30  & 2 \\
60--76   & 51 & 2 & 28 & 1 & 55  & 3  & 32  & 2 \\
76--86   & 97 & 4 & 56 & 3 & 98  & 5  & 58  & 3 \\
86--96   & 803& 14& 861& 15& 799 & 23 & 857 & 26\\
96--106  & 38 & 3 & 43 & 3 & 37  & 3  & 41  & 3 \\
106--120 & 12 & 1 & 12 & 1 & 11  & 1  & 12  & 1 \\
120--150 & 9.2 & 0.9& 9.7 & 1.0 & 8.4 & 0.8 & 8.8 & 0.9\\
150--200 & 3.1 & 0.6& 3.2 & 0.7 & 2.6 & 0.5 & 2.7 & 0.6\\
200--600 & 1.8 & 0.4& 1.9 & 0.5 & 1.4 & 0.3 & 1.5 & 0.4\\
\hline
\end{tabular}
\end{center}
\end{table}

\begin{table}[h!]
\begin{center}
\caption{Results for the DY spectrum normalized to the Z region in the
dielectron channel. The statistical and systematic uncertainties
are summed in quadrature.   $\RPOSTFSR$ and $\RPOSTFSRDET$ are
calculated using Eqs.~(\ref{eqn:fullCrossSectionRatio}) and~(\ref{eqn:fullCrossSectionRatio_DET}), respectively. The $\RDET$ and $R$ are
calculated using the FSR corrections.
\label{tab_result_electrons} }
\begin{tabular}{|l|r@{$~\pm~$}l|r@{$~\pm~$}l|r@{$~\pm~$}l|r@{$~\pm~$}l|}
\hline
Invariant mass bin (\!\GeV) 
& \multicolumn{2}{c|}{$\RPOSTFSRDET~(10^{-3})$} 
& \multicolumn{2}{c|}{$\RDET~(10^{-3}) $}
& \multicolumn{2}{c|}{$\RPOSTFSR~(10^{-3}) $}
& \multicolumn{2}{c|}{$R~(10^{-3}) $}\\
\hline
15--20  & 6   & 3  & 6   & 3  & 487 & 230 & 508 & 238\\
20--30  & 13  & 2  & 13  & 2  & 536 & 96  & 559 & 97\\
30--40  & 24  & 4  & 22  & 4  & 129 & 22  & 131 & 21\\
40--50  & 28  & 4  & 24  & 4  & 52  & 8   & 47  & 7\\
50--60  & 30  & 5  & 19  & 3  & 39  & 6   & 27  & 4\\
60--76  & 78  & 12 & 30  & 4  & 84  & 13  & 36  & 5\\
76--86  & 144 & 60 & 61  & 25 & 147 & 60  & 64  & 26\\
86--96  & 722 & 62 & 839 & 60 & 715 & 62  & 834 & 60\\
96--106 & 44  & 21 & 55  & 26 & 43  & 20  & 53  & 25\\
106--120& 13  & 3  & 15  & 3  & 12  & 2   & 14  & 3\\
120--150& 5.4 & 1.2& 6.0 & 1.3& 4.8 & 1.1 & 5.4 & 1.2  \\
150--200& 2.5 & 0.8& 2.8 & 0.8& 2.1 & 0.6 & 2.3 & 0.7  \\
200--600& 2.1 & 0.6& 2.4 & 0.7& 1.5 & 0.5 & 1.7 & 0.5  \\ \hline
\end{tabular}
\end{center}
\end{table}

The results are also normalized to the invariant mass bin widths, $\Delta M_i$, defining
\begin{equation}\label{eqn:shape_r}
   r_i = \frac{R_i}{\Delta M_i} .
   \end{equation}

Assuming lepton universality, the dimuon and dielectron results for $r_i$ are
combined in a weighted average, using as weights the inverse of the
respective squared total uncertainties, where the statistical and
systematic uncertainties are added in quadrature.

\begin{figure}[h!]
{\centering
\includegraphics[width=0.45\textwidth, angle=90]{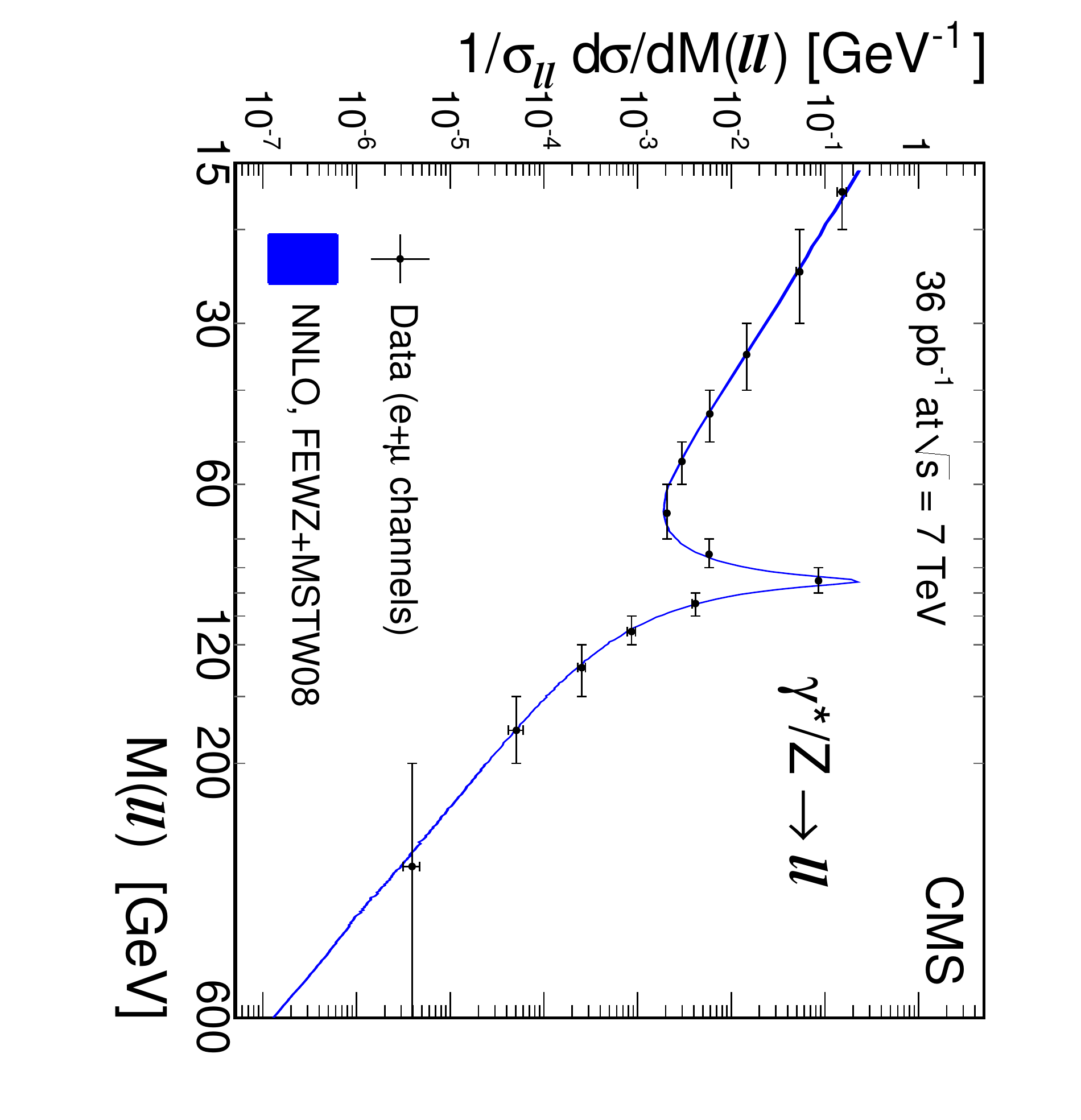}
\caption{\label{results}
DY invariant mass spectrum, normalized to the Z resonance region, 
$r = (1/\sigma_{\mathrm{\ell\ell}}) d\sigma/d\MLL$, as measured 
and as predicted by
NNLO calculations, for the full phase space\cite{DY_CMS}. The vertical error bar indicates 
the experimental (statistical and systematic) uncertainties summed in quadrature with
the theory uncertainty
resulting from the model-dependent kinematic distributions inside each bin.
The horizontal bars indicate bin sizes and the data points inside are placed according to  Ref.~\cite{binCorrection}. 
The width of the theory curve represents theoretical uncertainties which do not exceed few percent.}}
\end{figure}

Figure~\ref{results} compares the measured (combined)
results for the shape $r$ with the prediction from the FEWZ NNLO calculations, performed with the MSTW2008 PDF set \cite{ref:MSTW2008}. The measurements are very well reproduced by the theoretical calculations.

More details on the measurements can be found in Ref.~\cite{DY_CMS}.

\section{Conclusions}

The Drell--Yan differential cross section normalized to the cross section in the Z region
has been measured in pp collisions at $\sqrt{s} = 7$~\TeV,
in the dimuon and dielectron channels in the invariant mass range $15 < \MLL < 600$~\GeV.
The measurement is based on event samples collected by the CMS experiment,
corresponding to an integrated luminosity of $35.9\pm 1.4~{\mathrm{pb}}^{-1}$.
Results are presented both inside the detector acceptance and in the full phase space,
and the effect of final state QED radiation on the results is reported as well.
A correct description of the measurements requires modeling to NNLO
for dilepton invariant masses below about 30~\GeV.
The measurements are in good agreement with the NNLO theoretical predictions,
as computed with {\sc FEWZ}. 


\begin{acknowledgments}

We would like to thank the authors of {\sc FEWZ} and {\sc POWHEG} for the fruitful discussions, co-operation, and cross-checks in performing the theoretical calculations for our analysis.

\end{acknowledgments}

\bigskip 

\end{document}